\documentclass{ws-mpla}

\begin{document}

\renewcommand{\draftnote}{}
\renewcommand{\trimmarks}{}

\markboth{E. A. Matute} {Presymmetry with adulterated Dirac
neutrinos}

\catchline{}{}{}{}{}

\title{Presymmetry in the Standard Model with adulterated Dirac
neutrinos}

\author{\footnotesize Ernesto A. Matute}

\address{Departamento de F\'{\i}sica, Universidad de Santiago de
Chile,\\ Usach, Casilla 307 -- Correo 2, Santiago, Chile\\
ernesto.matute@usach.cl}

\maketitle

\pub{}{}

\begin{abstract}
Recently we proposed a model for light Dirac neutrinos in which
two right-handed (RH) neutrinos per generation are added to the
particles of the Standard Model (SM), implemented with the
symmetry of fermionic contents. The ordinary one is decoupled via
the high scale type-I seesaw mechanism, while the extra pairs off
with its left-handed (LH) partner. The symmetry of lepton and
quark contents was merely used as a guideline to the choice of
parameters because it is not a proper symmetry. Here we argue that
the underlying symmetry to take for this correspondence is
presymmetry, the hidden electroweak symmetry of the SM extended
with RH neutrinos defined by transformations which exchange lepton
and quark bare states with the same electroweak charges and no
Majorana mass terms in the underlying Lagrangian. It gives a
topological character to fractional charges, relates the number of
families to the number of quark colors, and now guarantees the
great disparity between the couplings of the two RH neutrinos.
Thus, Dirac neutrinos with extremely small masses appear as
natural predictions of presymmetry, satisfying the 't~Hooft's
naturalness conditions in the extended seesaw where the extra RH
neutrinos serve to adulterate the mass properties in the low scale
effective theory, which retains without extensions the gauge and
Higgs sectors of the SM. However, the high energy threshold for
the seesaw implies new physics to stabilize the quantum
corrections to the Higgs boson mass in agreement with the
naturalness requirement.

\end{abstract}

\ccode{PACS Nos.: 14.60.St, 14.60.Pq, 11.30.Hv, 11.30.Ly}

\section{Introduction}
\label{introduction}

The nonzero mass of neutrinos is one of the most compelling
evidences for physics beyond the Standard Model (SM) based on the
gauge groups $\mbox{SU}(3)_c \times \mbox{SU}(2)_L \times
\mbox{U}(1)_Y$ and Higgs fields in a single
doublet.\cite{SM1}$^{\mbox{--}}$\cite{SM3} In the SM, neutrinos do
not have Dirac mass as in the case of charged leptons and quarks
because only left-handed (LH) neutrinos are included. They do not
possess Majorana mass either, as $B-L$ is an (accidental) exact
global symmetry of the SM, where $B$ and $L$ denote the baryon and
lepton numbers.

In order to produce generic mass terms, one must then extend the
SM by adding three right-handed (RH) neutrinos to arrange Dirac
mass terms and also break the $B-L$ symmetry through RH Majorana
mass terms, which are allowed by the gauge symmetry of the SM. It
is a minimal expansion where the gauge and Higgs sectors of the SM
are maintained. Invoking the naturalness criterion of
't~Hooft,\cite{tHooft} this breakdown may be small since Majorana
mass terms equal to zero recover the $B-L$ symmetry and the model
becomes more symmetric. Actually, this is the well-known
pseudo-Dirac scenario,\cite{Wolf}$^{\mbox{--}}$\cite{Koba} where
the dominant contribution to neutrino masses comes from Dirac mass
terms with small corrections from Majorana mass terms. It appears
as an alternative to the popular seesaw
approach,\cite{Minkowski}$^{\mbox{--}}$\cite{Moha} where neutrino
mass terms are preponderantly of Majorana type, assuming a high
scale of new physics which leads to very different masses for
light LH and heavy RH neutrinos. This scenario also makes natural
in the sense of 't~Hooft the tiny mass of neutrinos since the
symmetry of lepton number is restored in the limit of small mass
terms equal to zero.

Thus far, the problem on the Dirac or Majorana nature of massive
neutrinos remains unresolved. As a matter of fact, no signals in
the search for neutrinoless double-beta decay of nuclei have been
observed,\cite{2beta1}$^{\mbox{--}}$\cite{2beta3} which is at
present the most feasible process capable of establishing the
Majorana nature of neutrinos. Hence, light neutrinos can be
Dirac-like fermions. In this work we restrict ourselves to this
plausible possibility, which is not so much explored in the
literature.

Within the framework of the SM just extended with three RH
neutrinos, there is no known natural mechanism to accounting for
the smallness of Dirac neutrino masses in comparison with the
charged lepton masses. The inclusion of small Majorana mass terms,
in addition to Dirac mass terms, as in the pseudo-Dirac scenario,
does not explain why neutrino masses are so light compared to
those of charged leptons. The inclusion of heavy Majorana mass
terms, as done in the type-I seesaw mechanism, only leads to light
Majorana neutrinos, but not to Dirac neutrinos. Hence, explaining
the Dirac nature of light neutrinos requires more physics, beyond
adding three RH neutrinos with small or large Majorana mass terms.
In other words, in the extension of the SM where a RH neutrino per
generation is introduced, the small value of the Dirac mass of
neutrinos compared to the charged leptons is not natural, in the
sense of the 't Hooft's principle of naturalness, so that a
fine-tuning is badly needed.

On the other hand, the current experimental status magnifies the
disturbing possibility that the successes of the SM may continue
at the TeV scale, so that the new runs of the Large Hadron
Collider (LHC) do not produce any significant hint of the new
physics which introduces extra gauge and Higgs fields with
breaking scales at the TeV range. If this simple view of the
experimental situation is adopted, one is left with the SM and the
presumed Dirac neutrinos with extremely small masses, for which
there is no explanation. This scenario can be expanded to
accommodate dark matter, for instance, by introducing extra
sterile neutrinos, although this point is not addressed here.

Even assuming this very conservative scenario, however, we
proposed for the first time in Ref.~\refcite{2RHne} a simple
approach to understanding the small masses of Dirac neutrinos in
comparison with charged leptons where masses are adulterated by
adding in each generation of the SM a second, almost inert, RH
neutrino with small Majorana mass. It is a minimal extension in
which the gauge and Higgs fields of the SM are kept, consistent
with the fact that the experimental evidence at the TeV scale may
still support the physics of the SM with massive neutrinos of
Dirac type. It applies the seesaw mechanism to suppress the
ordinary RH states with heavy Majorana masses and uses the
surviving extra RH neutrinos to generate the tiny Dirac masses,
i.e. the actual nature of light neutrinos would be of adulterated
Dirac type, where the usual RH neutrino is replaced by the extra
one of much smaller couplings. The key ingredient of this model to
making natural the large difference between the Majorana masses of
the two RH neutrinos in each generation is the correspondence
between lepton and quark contents when one of the RH neutrinos is
introduced in each generation. But this relation of particle
contents is not a proper symmetry to invoking the 't~Hooft's
principle of naturalness, i.e. there is no symmetry transformation
between leptons and quarks that maintains the Lagrangian in the
model invariant as they have different charges and Majorana mass
terms in the quark sector are absent. Therefore such a
correspondence was regarded just as a guideline to the choice of
parameters, conceding that the proper symmetry behind it should be
founded within the SM with RH neutrinos itself.

From another standpoint, the SM extended with three RH Dirac
neutrinos has been considered to re-establishing the electroweak
lepton--quark symmetry and incorporate
presymmetry,\cite{EAM1,EAM2} the symmetry hidden by the nontrivial
topology of the weak gauge fields which goes with the
lepton--quark symmetry from weak to electromagnetic interactions,
where a symmetry of lepton and quark contents is demanded. More
specifically, the symmetric electroweak patterns have been
explained in terms of underlying bare states of leptons and quarks
having the same charges and no Majorana mass terms, but located in
a topologically-nontrivial vacuum of the weak gauge fields in a
manner that the charge shifts are induced, in theory, via vacuum
tunneling weak instantons. Consequently, fractional charges get a
topological character and the number of families becomes
associated with the number of quark colors. And presymmetry
transformations exchange the bare states of leptons and quarks
keeping the underlying Lagrangian invariant.

Our aim in this work is to build up a consistent model for light
Dirac-type neutrinos, establishing presymmetry as the underlying
symmetry required to substantiate the symmetry of lepton and quark
contents used in the SM extended with two RH neutrinos per
generation, first discussed in Ref.~\refcite{2RHne}. This unifies
models of neutrinos and presymmetry, showing that Dirac-like
neutrinos with masses exceptionally small compared to charged
leptons are natural predictions of presymmetry, in the sense of
't~Hooft. As described above, all of these motivated by the
successes of the SM well above the TeV scale and the actual,
though not so well explored possibility that light neutrinos have
a Dirac nature.

Specifically, the motivation for presymmetry is the finding of a
proper symmetry which distinguishes the two RH neutrinos. Allowing
this concrete symmetry entails that the Dirac and Majorana mass
terms of the extra RH neutrinos are not free parameters,
independent of the original RH neutrinos. And they are not made
small by fine-tuning. Their smallness compared to the original RH
neutrinos are guaranteed by the presymmetry imposed on the theory
with these ones at the high-energy seesaw scale, much heavier than
the electroweak symmetry breaking scale. The 't~Hooft's argument
of naturalness for the small values of the Dirac and Majorana
masses of the extra RH neutrinos in the Lagrangian relies on this
presymmetry with the usual RH neutrinos; as the couplings of the
extra RH neutrinos tend to zero, the underlying theory only
involving the original RH neutrinos becomes more symmetric. This
symmetry guarantees the quantum corrections of such parameters to
be proportional to the parameters themselves and its interplay
with the seesaw mechanism leading to the low-energy effective
theory with the original RH neutrinos decoupled only introduces
omissible tiny corrections to the mass parameters. In particular,
the smallness of the Dirac mass with the extra RH neutrino in
comparison with the charged leptons appears robust.

Yet, there is no natural protection of the Higgs boson mass of the
SM against the large quantum corrections introduced by the high
scale of the seesaw for neutrino masses. The problem of
naturalness arises from the disparity between the energy scales
for the seesaw threshold and its upper value allowed by the
natural condition.\cite{Vissani} In order to maintain the
stability of the Higgs mass, the new physics associated with the
seesaw must then suppress the new contributions. The only best
known manner of having this cancellation in agreement with the
naturalness requirement is through supersymmetry. What is more, it
can be realized partially, as recently explored in
Ref.~\refcite{Urbano}, where the SM is considered
non-supersymmetric in the first stage. The implementation of our
extended seesaw with the new physics able to control the quantum
corrections to the Higgs mass, however, is beyond the scope of
this work.

The paper is organized as follows. In Sec.~\ref{mixed}, we recall
for completeness and subsequent discussion the needed results
about neutrino masses in the mixed scenario of Dirac and Majorana
neutrinos where two RH neutrinos per generation are added. In
Sec.~\ref{SM}, we look into the SM expanded with the adulterated
Dirac neutrinos of naturally small masses, where ordinary RH
neutrinos are decoupled. In Sec.~\ref{presym}, we go over
presymmetry in the extension of the SM with RH neutrinos and its
relevance in the model of light Dirac neutrinos. In
Sec.~\ref{limits}, we refer to phenomenological implications of
the unified model of massive neutrinos and presymmetry. The
conclusions are summarized in Sec.~\ref{conclusions}.

\section{Generic Neutrino Masses with Two Right-Handed Neutrinos
per Generation} \label{mixed}

The addition of RH neutrinos and the violation of lepton number
conservation are modifications of the SM in order to produce
neutrino masses in a generic way. Its expansion with two RH
neutrinos in each generation is to construct light Dirac neutrinos
with the extra one. In the following we review, for completeness
and subsequent discussion, the basic results for the SM extended
with two RH neutrinos per generation, preserving its gauge and
Higgs structures.\cite{2RHne} The first RH neutrino is the
ordinary one, which may carry a $B-L$ charge and form a doublet
with its RH charged lepton partner, as in models of left--right
symmetry.\cite{LRmodel1}$^{\mbox{--}}$\cite{LRmodel3} The other is
a secondary singlet with generally small couplings and no local
charges. The crucial element of the model is the symmetry of
lepton and quark contents when just one of the RH neutrinos is
added. In this manner, invoking the 't~Hooft's
criterion,\cite{tHooft} the smallness of couplings of the second
RH neutrino appears natural since the symmetry of quark and lepton
contents with the first RH neutrino is re-established if these
couplings are set to be zero.

The symmetry of fermionic contents, however, is merely used as a
guideline to the choice of parameters because it is not a proper
symmetry in the Lagrangian. This means that one cannot define a
symmetry transformation between leptons and quarks to keep the
Lagrangian invariant, as these particles have different charges
and Majorana mass terms are not present in the quark sector. In
Sec.~\ref{presym}, we discuss on the proper symmetry that must be
attached to this symmetry of lepton and quark contents when one of
the RH neutrinos is introduced for each generation.

The mass terms after spontaneous electroweak symmetry breaking
are\cite{2RHne}
\begin{equation}
- {\cal L}_\nu = \frac{1}{2} \left( \begin{array}{ccc} \bar{\nu}_L
& \bar{\nu}^c_L & \bar{\nu}^{\prime c}_L \end{array} \right)
\left(
\begin{array}{ccc}
0 & M_D & M^\prime_D \\ M^T_D & M_R & M^{\prime T} \\
M^{\prime T}_D & M^\prime & M^\prime_R \end{array} \right) \left(
\begin{array}{c} \nu^c_R \\ \nu_R \\ \nu^\prime_R
\end{array} \right) + \; h.c. ,
\label{massL}
\end{equation}
where $\nu_R$ ($\nu^\prime_R$) is a three-component vector
denoting the ordinary (extra) RH neutrinos, and $M_D$,
$M^\prime_D$ are 3$\times$3 complex matrices referring to the
Dirac mass terms, $M_R$, $M^\prime_R$ to the RH Majorana mass
terms, and $M^\prime$ to the mixing terms; the phase convention is
$\nu^c_R={\mathcal C}\bar{\nu}^T_L$.

The values of masses and couplings of RH neutrinos should be
understood. Because they are not part of the SM, whose origin in
turn is not known yet, the form of their findings may not be well
defined. In the SM every LH charged lepton and quark has its RH
charged lepton or quark partner, while the RH partner of the
neutrino is absent. This content of chiral fermions is clearly not
symmetric. The simplest manner of having such a symmetry between
leptons and quarks is through the introduction of ordinary RH
neutrinos, $\nu_R$.\cite{Shapo,Branco} Within the formalism of
Eq.~(\ref{massL}), it corresponds to $M^\prime_D=0$, $M^\prime=0$
and $M^\prime_R=0$, but keeping $M_D$ and $M_R$ as nontrivial.
Here the proposal embraces the rationale of the type-I seesaw
mechanism based on the assumptions that $M_D$ has the same mass
scale as charged leptons, and $M_R$ is sufficiently large to
suppress $M_D$ according to $M_D M^{-1}_R M^T_D$. Thus, the
lepton--quark symmetry of particle content together with a large
$M_R$ and $M^\prime_D = M^\prime = M^\prime_R = 0$ mimic a high
scale type-I seesaw, decoupling the ordinary RH neutrinos.

The lepton--quark correspondence, however, is broken when the
extra RH neutrinos, $\nu^\prime_R$, are added. This is regarded as
a reason for having small couplings $M^\prime_D$, $M^\prime$,
$M^\prime_R$ for $\nu^\prime_R$, as the 't~Hooft's naturalness
criterion applied to this symmetry of lepton and quark contents in
the Lagrangian gives a ready explanation. Indeed these extra RH
neutrinos establish an alternative, exceptionally weak,
lepton--quark correspondence. Therefore the symmetry of fermionic
contents distinguishes $\nu_R$ from $\nu^\prime_R$ by requiring a
large difference between $M_D$, $M_R$ and $M^\prime_D$,
$M^\prime_R$, respectively, which parametrize the two forms of the
symmetry. This other lepton--quark symmetry of particle content
together with $M_D = M^\prime = M_R = 0$ and small $M^\prime_R$
mimic a low scale pseudo-Dirac scenario, but pairing off the LH
with the extra RH neutrinos. As emphasized above, the
correspondence of leptons and quarks is only considered as a
guideline to coupling selections since it is not a proper symmetry
in the electroweak Lagrangian.

The mass matrix in Eq.~(\ref{massL}) is diagonalizable by the
unitary transformation
\begin{equation}
{\cal U}^\dag {\cal M} {\cal U}^{*} = \left(
\begin{array}{ccc} D_L & 0 & 0 \\ 0 & D_R & 0 \\ 0 & 0 &
D^\prime_R \end{array} \right) ,
\end{equation}
where $D_L$, $D_R$, and $D^\prime_R$ are diagonal, real, and
non-negative 3$\times$3 matrices. The unitary matrix can be
written as
\begin{equation}
{\cal U}^\dagger = \left( \begin{array}{ccc} V^\dagger_L & 0 & 0
\\ 0 & V^\dagger_R & 0 \\ 0 & 0 & V^{\prime \dagger}_R \end{array}
\right) \left(
\begin{array}{ccc}
(\frac{1}{\sqrt{2}}I+W^\dagger_{LL}) \;\; & V^\dagger_{RL} \;\; &
(-\frac{1}{\sqrt{2}}I+W^{\prime \dagger}_{RL}) \\
V^\dagger_{LR} & I & V^{\prime \dagger}_{RL} \\
(\frac{1}{\sqrt{2}}I+W^{\prime \dagger}_{LR}) & V^{\prime
\dagger}_{LR} & (\frac{1}{\sqrt{2}}I+W^{\prime \dagger}_{RR})
\end{array} \right) ,
\end{equation}
where $V_L$, $V_R$, and $V^\prime_R$ are unitary 3$\times$3
complex matrices. If it is assumed that $M_R$ and $M^\prime_D$ are
non-singular and symmetric matrices, and that $M^\prime_R$,
$M^\prime$, $M^\prime_D$, $M_D \ll M_R$ and $M^\prime_R$, $M_D
M^{-1}_R M^T_D$, $M^\prime M^{-1}_R M^{\prime T}$, $M^\prime
M^{-1}_R M^T_D \ll M^\prime_D$, as argued above, and the
constraints from unitarity and the matrix ${\cal MU}^{*}$ are used
as in the ordinary seesaw mechanism, the following expressions are
obtained:
\begin{eqnarray}
\begin{array}{l}
W^\dagger_{LL} \simeq \frac{1}{4\sqrt{2}} M^\prime_R M^{\prime
-1}_D + \frac{1}{4\sqrt{2}} (M_D-M^\prime) M^{-1}_R
(M^T_D+M^{\prime T}) M^{\prime -1}_D , \\ [10pt] W^{\prime
\dagger}_{RR} \simeq \frac{1}{4\sqrt{2}} M^\prime_R M^{\prime
-1}_D + \frac{1}{4\sqrt{2}} (M_D+M^\prime) M^{-1}_R
(M^T_D-M^{\prime T}) M^{\prime -1}_D , \\ [10pt] W^{\prime
\dagger}_{RL} \simeq W^\dagger_{LL} , \\ [10pt] W^{\prime
\dagger}_{LR} \simeq - W^{\prime \dagger}_{RR} , \\ [10pt]
V^\dagger_{RL} \simeq - (\frac{1}{\sqrt{2}}I+W^\dagger_{LL}) M_D
M^{-1}_R + (\frac{1}{\sqrt{2}}I-W^\dagger_{LL}) M^\prime M^{-1}_R
, \\ [10pt]  V^{\prime \dagger}_{LR} \simeq -
(\frac{1}{\sqrt{2}}I-W^{\prime \dagger}_{RR}) M_D M^{-1}_R -
(\frac{1}{\sqrt{2}}I+W^{\prime \dagger}_{RR}) M^\prime M^{-1}_R ,
\\ [10pt] V^\dagger_{LR} \simeq M^{-1 \dagger}_R M^\dagger_D
, \\ [10pt] V^{\prime \dagger}_{RL} \simeq M^{-1 \dagger}_R
M^{\prime \dagger} .
\end{array}
\label{Ws}
\end{eqnarray}
These lead to
\begin{eqnarray}
D_L &\simeq& V^\dagger_L [-M^\prime_D+\frac{1}{2} M^\prime_R -
\frac{1}{2} (M_D-M^\prime)M^{-1}_R(M^T_D-M^{\prime T})] V^{*}_L
\nonumber \\ &\simeq& - V^\dagger_L M^\prime_D V^{*}_L , \nonumber \\
D_R &\simeq& V^\dagger_R M_R V^{*}_R , \label{Ds} \\
D^\prime_R &\simeq& V^{\prime \dagger}_R [M^\prime_D+\frac{1}{2}
M^\prime_R - \frac{1}{2} (M_D+M^\prime)M^{-1}_R(M^T_D+M^{\prime
T})] V^{\prime *}_R \nonumber \\ &\simeq& V^{\prime \dagger}_R
M^\prime_D V^{\prime *}_R . \nonumber
\end{eqnarray}

In the mixed pseudo-Dirac and seesaw regimes with $M^\prime_R=0$
and $M_D$, $M^\prime$ kept down, there are three light almost
degenerate pairs of mass eigenstates with small mass differences,
having almost maximal mixing of LH neutrinos $\nu_L$ and
adulterant RH neutrinos $\nu^\prime_R$, and three heavy, mostly
ordinary RH neutrinos $\nu_R$ with mass matrix $M_R$. The mass of
light neutrinos are of the order of $M^\prime_D$ instead of $M_D$,
which are down by the seesaw mechanism. The incidence of matrices
$V_{LR}$, $V_{RL}$, $V^\prime_{LR}$, and $V^\prime_{RL}$ are
reduced by $M_R$, while $W_{LL}$, $W^\prime_{RR}$,
$W^\prime_{LR}$, and $W^\prime_{RL}$ are by $M_R$ and/or
$M^\prime_D$.

\section{The Standard Model with Adulterated Dirac Neutrinos}
\label{SM}

The RH neutrinos with large masses can be integrated out by means
of the equation of motion
\begin{equation}
\frac{d {\cal L}_\nu}{d \nu_R} = 0 , \label{motionEq}
\end{equation}
which gives
\begin{equation}
\bar{\nu}^c_L = - \bar{\nu}_L M_D M^{-1}_R - \bar{\nu}^{\prime
c}_L M^\prime M^{-1}_R , \hspace{20pt} \nu_R = - M^{-1}_R M^T_D
\nu^c_R - M^{-1}_R M^{\prime T} \nu^\prime_R .
\end{equation}
The effective Lagrangian is then written as
\begin{equation}
- {\cal L}_\nu = \frac{1}{2} \left( \begin{array}{cc} \bar{\nu}_L
& \bar{\nu}^{\prime c}_L \end{array} \right) \left(
\begin{array}{cc}
M_{LL} \; & M^\prime_{LR} \\ M^{\prime T}_{LR} & M^\prime_{RR}
\end{array} \right)
\left( \begin{array}{c} \nu^c_R \\ \nu^\prime_R
\end{array} \right) + h.c. ,
\label{effL}
\end{equation}
where
\begin{eqnarray}
\begin{array}{c}
M_{LL} \simeq - M_D M^{-1}_R M^T_D , \qquad M^\prime_{LR} \simeq
M^\prime_D - M_D M^{-1}_R M^{\prime T} , \\ [10pt] M^\prime_{RR}
\simeq - M^\prime M^{-1}_R M^{\prime T} .
\end{array}
\label{Masses}
\end{eqnarray}

The mass matrix in Eq.~(\ref{effL}) is diagonalized by the
approximately unitary matrix
\begin{equation}
{\cal U}^\dagger \simeq \left( \begin{array}{cc} V^\dagger_L & 0
\\ & \\ 0 & V^{\prime \dagger}_R \end{array} \right) \left(
\begin{array}{cc}
(\frac{1}{\sqrt{2}}I+W^\dagger_{LL}) \; &
(-\frac{1}{\sqrt{2}}I+W^\dagger_{LL}) \\ & \\
(\frac{1}{\sqrt{2}}I-W^{\prime \dagger}_{RR}) &
(\frac{1}{\sqrt{2}}I+W^{\prime \dagger}_{RR})
\end{array} \right) ,
\end{equation}
such that
\begin{equation}
{\cal U}^\dag {\cal M} {\cal U}^{*} = \left(
\begin{array}{cc} D_L & 0 \\ 0 & D^\prime_R \end{array}
\right) ,
\end{equation}
where $W^\dagger_{LL}$ and $W^{\prime \dagger}_{RR}$, $D_L$ and
$D^\prime_R$ have the expressions given in Eqs.~(\ref{Ws}) and
(\ref{Ds}) with $M^\prime_R=0$.

Now, the mass hierarchies
\begin{equation}
M_D M^{-1}_R M^T_D , \, M_D M^{-1}_R M^{\prime T} , \, M^\prime
M^{-1}_R M^{\prime T} \ll M^\prime_D \ll M_D \ll M_R ,
\label{hierarchy}
\end{equation}
lead to the mass matrix
\begin{equation}
{\cal M} \simeq \left(
\begin{array}{cc} 0 & M^\prime_D \\ M^{\prime T}_D & 0
\end{array} \right) ,
\end{equation}
which is consistent with the SM extended with the extra RH
neutrinos, $\nu^\prime_R$, having a Dirac nature. Actually, in the
vanishing limit of the small values $M_D M^{-1}_R M^T_D$, $M_D
M^{-1}_R M^{\prime T}$, $M^\prime M^{-1}_R M^{\prime T}$, a lepton
number conservation and a lepton--quark symmetry are set up at low
energies. It is the adulterated lepton--quark symmetry of particle
content in terms of $\nu_R^\prime$ with all couplings of $\nu_R$
removed ($M_D=M_R=M^\prime=0$). Dirac neutrino masses much smaller
than those of charged leptons now appear natural because
$M^\prime_D=0$ (with $M^\prime=M^\prime_R=0$) restores an enhanced
symmetry in the original Lagrangian, specifically, the symmetry of
lepton and quark contents including the ordinary neutrino partners
$\nu_R$. Thus adulterated Dirac neutrinos with tiny masses can be
accommodated naturally. Once more, we emphasize that the
correspondence between lepton and quark contents just serves as a
guideline to the choice of parameters since it is not a proper
symmetry in the electroweak Lagrangian.

\section{Presymmetry in the Standard Model with Right-Handed
Neutrinos} \label{presym}

We now substantiate the lepton--quark symmetry of particle content
through a proper symmetry of the SM extended with RH neutrinos,
based on the gauge groups $\mbox{SU}(3)_c \times \mbox{SU}(2)_L
\times \mbox{U}(1)_Y$ and Higgs fields in a doublet. An available
simple option for this symmetry is presymmetry, where the crucial
elements to have a well-defined symmetry transformation appear
naturally. In fact, presymmetry is conceived as a symmetry of an
electroweak Lagrangian under transformations on underlying bare
states of leptons and quarks having the same charges and no
associated Majorana mass terms. To make the whole in a consistent
way, we review in the following the key arguments for presymmetry
when one RH neutrino is added in each generation.

On the one hand, there is the following hypercharge symmetry
between chiral leptons and quarks within each of their three
families:\cite{EAM1,EAM2}
\begin{eqnarray}
\begin{array}{ll}
Y(\nu_{L}) = Y(u_L) + \Delta Y(u_L) = -1 \, , \qquad & Y(e_L) =
Y(d_L) + \Delta Y(d_L) = -1 \, , \\ [10pt] Y(\nu_{R}) = Y(u_R) +
\Delta Y(u_R) = 0 \, , \qquad & Y(e_R) = Y(d_R) + \Delta Y(d_R) =
-2 \, ,
\end{array}
\label{Ysym1}
\end{eqnarray}
and, on the other hand,
\begin{eqnarray}
\begin{array}{ll}
\displaystyle Y(u_L) = Y(\nu_{L}) + \Delta Y(\nu_{L}) =
\frac{1}{3} \, , \qquad & \displaystyle Y(d_L) = Y(e_L) + \Delta
Y(e_L) = \frac{1}{3} \, , \\ [10pt] \displaystyle Y(u_R) =
Y(\nu_{R}) + \Delta Y(\nu_{R}) = \frac{4}{3} \, , \qquad &
\displaystyle Y(d_R) = Y(e_R) + \Delta Y(e_R) = - \frac{2}{3} \, ,
\end{array}
\label{Ysym2}
\end{eqnarray}
with the $\Delta Y$ equal to $4/3$ for leptons and $-4/3$ for
quarks being related to the lepton and baryon numbers according to
\begin{equation}
\Delta Y = \frac{4}{3} \; (L-3B) \, ,
\label{DeltaY}
\end{equation}
where the conventional relation $Q=T_3+Y/2$ between electric
charge, weak isospin, and hypercharge, is used. Any other
hypercharge normalization can modify the value of the global
fractional piece $\Delta Y$, but the charge symmetry described in
Eqs.~(\ref{Ysym1}) and (\ref{Ysym2}) is preserved.\cite{EAM3}

Presymmetry is associated with the equality of lepton and quark
charges when the global part $\Delta Y$ is set apart. The
inclusion of RH neutrinos is essential to completing this
correspondence between charges, which in turn allows a symmetry of
lepton and quark contents when one RH neutrino per generation is
introduced; the symmetric pattern in terms of the extra,
adulterant RH neutrinos considers $\nu^\prime_{R}$ instead of
$\nu_{R}$ in Eqs.~(\ref{Ysym1}) and (\ref{Ysym2}).

The charge symmetry and charge dequantization underlying
Eqs.~(\ref{Ysym1}) and (\ref{Ysym2}) can be understood if
prelepton and prequark states are taken into account. These are
defined by the quantum numbers of leptons and quarks,
respectively, except charge values, and denoted by a hat accent
over the corresponding flavor symbol. The hypercharges of
preleptons and prequarks are the same as their respective quark
and lepton weak partners. From Eqs.~(\ref{Ysym1}) and
(\ref{Ysym2}) one is then led, in the first case,
to\cite{EAM1,EAM2}
\begin{eqnarray}
\begin{array}{ll}
Y(\nu_{L}) = Y(\hat{\nu}_{L}) + \Delta Y(\hat{\nu}_{L}) \, ,
\qquad & Y(e_L) = Y(\hat{e}_L) + \Delta Y(\hat{e}_L) \, ,
\\ [10pt] Y(\nu_{R}) = Y(\hat{\nu}_{R}) + \Delta
Y(\hat{\nu}_{R}) \, , \qquad & Y(e_R) = Y(\hat{e}_R) + \Delta
Y(\hat{e}_R) \, ,
\end{array}
\label{Ysym3}
\end{eqnarray}
and, in the other case, to
\begin{eqnarray}
\begin{array}{ll}
Y(u_L) = Y(\hat{u}_L) + \Delta Y(\hat{u}_L) \, , \qquad & Y(d_L) =
Y(\hat{d}_L) + \Delta Y(\hat{d}_L) \, , \\ [10pt] Y(u_R) =
Y(\hat{u}_R) + \Delta Y(\hat{u}_R) \, , \qquad & Y(d_R) =
Y(\hat{d}_R) + \Delta Y(\hat{d}_R) \, ,
\end{array}
\label{Ysym4}
\end{eqnarray}
with prelepton--quark charge symmetry established as
\begin{eqnarray}
\begin{array}{ll} Y(\hat{\nu}_{L}) = Y(u_L) \, , \qquad &
\Delta Y(\hat{\nu}_{L}) = \Delta Y(u_L) \, , \\ [10pt]
Y(\hat{\nu}_{R}) = Y(u_R) \, , \qquad & \Delta Y(\hat{\nu}_{R}) =
\Delta Y(u_R) \, , \\ [10pt] Y(\hat{e}_L)
= Y(d_L) \, , \qquad & \Delta Y(\hat{e}_L) = \Delta Y(d_L) \, , \\
[10pt] Y(\hat{e}_R) = Y(d_R) \, , \qquad & \Delta Y(\hat{e}_R) =
\Delta Y(d_L) \, ,
\end{array}
\label{lqPresym1}
\end{eqnarray}
and prequark--lepton charge symmetry given by
\begin{eqnarray}
\begin{array}{ll}
Y(\hat{u}_L) = Y(\nu_{L}) \, , \qquad & \Delta Y(\hat{u}_L) =
\Delta Y(\nu_{L}) \, , \\ [10pt] Y(\hat{u}_R) = Y(\nu_{R}) \, ,
\qquad & \Delta Y(\hat{u}_R) = \Delta Y(\nu_{R}) \, ,
\\ [10pt] Y(\hat{d}_L) = Y(e_L) \, , \qquad & \Delta Y(\hat{d}_L)
= \Delta Y(e_L) \, , \\ [10pt] Y(\hat{d}_R) = Y(e_R) \, , \qquad &
\Delta Y(\hat{d}_R) = \Delta Y(e_L) \, ,
\end{array}
\label{lqPresym2}
\end{eqnarray}
where the relation of $\Delta Y$ with the lepton and baryon
numbers for preleptons and prequarks is now
\begin{equation}
\Delta Y = \frac{4}{3} \; (3L-B) \, ,
\label{Delta2}
\end{equation}
with $\Delta Y$ equal to $-4/3$ for preleptons and $4/3$ for
prequarks. Thus, $L=-1/3$ for preleptons, with the $3$ being
attributed to the number of families, and $B=-1$ for
prequarks.\cite{EAM2}

The underlying lepton--quark charge symmetry when just one RH
neutrino is introduced for each generation has been set forth with
the global piece of hypercharge having a weak topological
character. It has been argued that the fact that any weak
topological property cannot have observable effects at the
zero-temperature scale because of the smallness of the weak
coupling, implies that the charge structures reflected in
Eqs.~(\ref{Ysym3}) and (\ref{Ysym4}) do not apply to physical
leptons and quarks, but to new underlying states referred to as
bare leptons and quarks which have topological
ingredients.\cite{EAM2} However, the assignments of these bare
leptons and quarks to the gauge groups of the SM are the same of
standard leptons and quarks. The electroweak presymmetry is
therefore between preleptons and bare quarks, and between
prequarks and bare leptons. Due to their topological properties,
preleptons and bare quarks have also been named topological
preleptons and topological quarks, respectively.

The interactions of topological preleptons and topological quarks,
as well as of prequarks and bare leptons, with the gauge fields
are supposed to be defined by the same Lagrangian of the gauge
sector of the SM with leptons and quarks excepting hypercharge
couplings. In the Yukawa sector, Majorana mass terms are forbidden
for RH preneutrinos since these have nonzero hypercharge, but they
are a possibility at the physical lepton--quark level. Presymmetry
is the invariance of the bare electroweak Lagrangian under flavor
transformations of a $Z_2$ group which exchange topological
preleptons and topological quarks on the one hand,
$\hat{\nu}_{L(R)} \leftrightarrow {u}_{L(R)}$, $\hat{e}_{L(R)}
\leftrightarrow {d}_{L(R)}$, and prequarks and bare leptons on the
other hand, $\hat{u}_{L(R)} \leftrightarrow {\nu}_{L(R)}$,
$\hat{d}_{L(R)} \leftrightarrow {e}_{L(R)}$.

The charge shifts are originated by the nonstandard hypercharges
of the new fermionic states, which produce gauge anomalies in the
couplings by fermion triangle loops of three currents related to
the chiral $\mbox{U(1)}_Y$ and $\mbox{SU(2)}_L$ gauge symmetries.
In fact, in the scenario of topological preleptons and quarks, for
example, the $\mbox{U(1)}_Y$ gauge current in all representations
\begin{equation}
\hat{J}^{\mu}_{Y} = \overline{\hat{\ell}}_{L} \gamma^{\mu}
\frac{Y}{2} \hat{\ell}_{L} + \overline{\hat{\ell}}_{R}
\gamma^{\mu} \frac{Y}{2} \hat{\ell}_{R} + \overline{q}_{L}
\gamma^{\mu} \frac{Y}{2} q_{L} + \overline{q}_{R} \gamma^{\mu}
\frac{Y}{2} q_{R} \, ,
\label{gaugecurrent}
\end{equation}
exhibits the $\mbox{U(1)}_Y[\mbox{SU(2)}_L]^2$ and
$[\mbox{U(1)}_Y]^3$ anomalies due to the nonvanishing of the
following sums which include one RH preneutrino per generation:
\begin{equation}
\sum_{L} Y = 8 \, , \qquad \sum_{LR} Y^{3} = - 24 \, ,
\end{equation}
where the first runs over the LH and the second over the LH and RH
topological preleptons and quarks, with $(-1)$ for the RH
contributions. Their cancellations need a counterterm which
contains topological currents or Chern--Simons classes associated
with the $\mbox{U(1)}_Y$ and $\mbox{SU(2)}_L$ gauge groups:
\begin{equation}
J^{\mu}_{T} = \frac{1}{4} \, K^{\mu} \sum_{L} Y + \frac{1}{16} \,
L^{\mu} \, \sum_{LR} Y^{3} = 2 \, K^{\mu} - \frac{3}{2} \, L^{\mu}
\, , \label{currX}
\end{equation}
where
\begin{eqnarray}
\begin{array}{rcl}
\displaystyle K^{\mu} &=& \displaystyle \frac{g^{2}}{8 \pi^{2}} \,
\epsilon^{\mu\nu\lambda\rho} \; \mbox{tr} \!\! \left( W_{\nu}
\partial_{\lambda} W_{\rho} - \frac{2}{3} \, i g
W_{\nu} W_{\lambda} W_{\rho} \right) , \\[12pt]
\displaystyle L^{\mu} &=& \displaystyle \frac{{g'}^{2}}{12
\pi^{2}} \, \epsilon^{\mu\nu\lambda\rho} A_{\nu}
\partial_{\lambda} A_{\rho} \, ,
\end{array}
\label{CS}
\end{eqnarray}
so that the new current $J^{\mu}_{Y} = \hat{J}^{\mu}_{Y} +
J^{\mu}_{T}$ is anomaly free, gauge noninvariant, and also
symmetric under the exchange of topological preleptons and quarks.
Moreover, its charge is not conserved due to the topological
charge which gives the change in the topological winding number of
the asymptotic, pure gauge field configurations, assuming that the
space--time region of nonzero energy density is bounded. Indeed,
advocating the principle of equality for all preleptons of the
system in the partition of the topological charge,\cite{EAM1} the
change in each charge, using Eqs.~(\ref{currX}) and (\ref{CS}) for
the pure gauge fields, is
\begin{equation}
\Delta Q_{Y} = \displaystyle \frac{1}{6} \, ( n_{+} - n_{-} ) =
\frac{1}{6} \, n \, , \label{deltaQ}
\end{equation}
with the topological charge given by
\begin{equation}
n = \int d^{4}x \, \partial_{\mu} K^{\mu} = \frac{g^{2}}{16
\pi^{2}} \int d^{4}x \, \mbox{tr} (W_{\mu\nu} \tilde{W}^{\mu\nu})
\, .
\label{QT1}
\end{equation}
These topological numbers vanish in the $\mbox{U(1)}_Y$ case.

Vacuum states labeled by different topological numbers are then
tunneled by $\mbox{SU(2)}_L$ instantons carrying topological
charges, making possible in principle the charge shifts and
transitions from fermions with nonstandard to those with standard
charges. Each hypercharge is changed by a same amount:
\begin{equation}
Y(\hat{\ell}) \rightarrow Y(\hat{\ell}) + \frac{n}{3} \, .
\label{norma}
\end{equation}
The value $n=-4$ leading to Eq.~(\ref{Delta2}) is set by the
cancellation of anomalies and elimination of the associated
counterterm (see Eq.~(\ref{currX})), demanded by the gauge
invariance and renormalizability of the theory; the coefficient
$3L-B$ in Eq.~(\ref{Delta2}) is just a counting number.\cite{EAM2}

According to the presymmetry model, topological preleptons have a
vacuum gauge field configuration of winding number $n_{-}=4$, if
gauge freedom is used to set $n_{+}=0$ for that of leptons. The
transformation of topological preleptons into bare leptons is via
a Euclidean topological weak instanton with topological charge
$n=-4$, conceived in Minkowski space--time as a quantum mechanical
tunneling event between vacuum states of weak $\mbox{SU(2)}_L$
gauge fields with different topological winding numbers. Thus,
topological preleptons and bare leptons are differentiated by the
topological vacua of their weak gauge configurations, tunneled by
a weak four-instanton bearing the topological charge and inducing
the global fractional piece of charge required for normalization.

However, the transitions from topological preleptons to bare
leptons by means of the weak $\mbox{SU(2)}_L$ instantons, as well
as those from prequarks to topological quarks, do not occur in the
actual world because topological preleptons, prequarks and
topological quarks are not physical dynamical entities. They are
bare states of leptons and quarks considered as mathematical
entities out of which the observed particle states are
constructed. In a sense, such transformations are frustrated by
the extreme smallness of the instanton transition probability at
zero temperature, and the charge normalization eliminates the
extraordinarily large time scale for them, leading to leptons and
quarks with trivial topology and standard charges. The replacement
of bare leptons and quarks with normalized charges by the standard
ones in the electroweak Lagrangian is straightforward as they have
the same quantum numbers,\cite{EAM1,EAM2} and the insertion of
Majorana mass terms for RH neutrinos gives the shape of the
extended effective theory.

Despite this, one still has a proper symmetry transformation at
the level of preleptons defined by the exchange of all topological
preleptons and quarks in the electroweak sector of the Lagrangian,
which requires a correspondence between fermionic contents at the
stages of preleptons and leptons. But this is precisely what is
needed to have a natural framework which allows light Dirac-like
neutrinos in the low-energy theory. In fact, as seen in
Secs.~\ref{mixed} and \ref{SM}, the addition of a second RH
neutrino per generation coupling \`{a} la Dirac to the LH neutrino
provides the seed for that as the relation between light neutrino
masses and the charged lepton masses is broken. By means of this,
the naturalness problem is solved, i.e. the question why light
Dirac neutrinos are so much lighter than charged leptons is
answered.

The well-defined presymmetry now validates the sequence of
hierarchies assumed in Secs.~\ref{mixed} and \ref{SM}. The first
hierarchy, $M_D \ll M_R$, mimics the standard high scale type-I
seesaw scenario in which only one RH neutrino per generation,
$\nu_R$, is introduced. The symmetry of lepton and quark contents
and the hypercharge symmetry in Eq.~(\ref{Ysym1}) are
re-established. At the underlying level of preleptons defined in
Eqs.~(\ref{Ysym3}) and (\ref{lqPresym1}), one has the presymmetry
transformations that exchange topological prelepton and quark
fields in a Lagrangian where Majorana mass terms are absent since
all preleptons have electroweak charges.

By introducing the second RH neutrino for each generation of
leptons and quarks, which breaks presymmetry, the second
hierarchy, $M^\prime_D \ll M_D$, is understood naturally since if
this new parameter is set equal to zero, the presymmetry involving
the first RH neutrino is recovered.

The last hierarchies, ($M_D M^{-1}_R M^T_D$, $M^\prime M^{-1}_R
M^{\prime T}$, $M_D M^{-1}_R M^{\prime T}$) $\ll$ $M^\prime_D$
(entailing $M_D,M^\prime$ $\ll$ $M_R$) and $M^\prime_R \ll
M^\prime_D$, mimic the low scale pseudo-Dirac scenario. If the
small ratios of mass parameters are neglected, for all practical
purposes the first RH neutrino is decoupled from the low energy
model. As such, the symmetry of lepton and quark contents, the
symmetric charge relations in Eq.~(\ref{Ysym1}) and in
Eqs.~(\ref{Ysym3}) and (\ref{lqPresym1}) at the underlying level,
as well as presymmetry, now engage $\nu^\prime_R$ instead of
$\nu_R$. In this way, the residual presymmetry connecting
$\nu^\prime_R$ makes the low scale model more symmetric. Thus all
the hierarchies are natural in the sense of t'~Hooft.

\section{Phenomenological Implications of the Model}
\label{limits}

In order to have light neutrinos of adulterated Dirac type, our
extension of the SM assumes two RH neutrinos in each generation
and the hierarchy of masses given in Eq.~(\ref{hierarchy}). As
discussed above, the first choice is to decouple through the
extended seesaw mechanism the original RH neutrinos from the
others by making them much heavier than the other mass parameters
and ratios of mass parameters. In a second step, the light
neutrino masses are effectively controlled by the new Dirac mass
of the extra RH neutrinos. This requires a hierarchy between the
extra Dirac masses and the other mass parameters and ratios of
mass parameters. Only this second hierarchy gives rise to the
light pseudo-Dirac neutrinos.

The Dirac mass $M_D$ is supposed to be of order the charged lepton
mass. In the approximation of one generation, if a mass $M_D =
{\cal O}(1 \, \mbox{MeV})$ is for the charged lepton of first
generation and the upper bounds
\begin{equation}
M_{LL} , M^\prime_{RR} \leq {\cal O}(10^{-9} \mbox{eV})
\end{equation}
obtained from data analyzes on solar neutrino
oscillations\cite{mRR,whitepaper} are considered, then
Eq.~(\ref{Masses}) leads to an energy threshold for the type-I
seesaw equal to $M_{R} = {\cal O}(10^{12} \, \mbox{GeV})$.
Besides, using the experimental data on neutrino mass,\cite{PDG}
\begin{equation}
M^\prime_{LR} = {\cal O}(10^{-1} \, \mbox{eV}) ,
\end{equation}
the indicative values for the parameters in the model become
\begin{eqnarray}
\begin{array}{c}
M_R \geq {\cal O} (10^{12} \, \mbox{GeV}) , \qquad M_D = {\cal O}
(1
\, \mbox{MeV}) , \qquad M^\prime_D = {\cal O} (10^{-1} \, \mbox{eV}) , \\
[10pt] M^\prime_R \leq {\cal O} (10^{-9} \, \mbox{eV}) , \qquad
M^\prime \leq {\cal O} (1 \, \mbox{MeV}) ,
\end{array}
\end{eqnarray}
which are consistent with the hierarchy of masses
\begin{equation}
M_{LL} , M^\prime_{RR} \ll M^\prime_{LR} \ll M_D \ll M_R
\end{equation}
adopted in the model, so ratifying the Dirac type assumed for
light neutrinos.

Thus the parameter region being considered excludes the
pseudo-Dirac limit, but not the Dirac character for light
neutrinos. Their masses or Yukawa couplings may have exceptionally
small values because of the adulterant character of RH partners.
Also, there is consistency between this Dirac picture and the
vanishing of the Majorana mass $M^\prime_R$ assumed above.
Besides, the Dirac nature of lighter neutrinos forbids the
neutrinoless double-beta decay, in accordance with recent
precision experiments.\cite{2beta1}$^{\mbox{--}}$\cite{2beta3}
Even more, no significant departures from the SM predictions are
expected at the TeV region, leading to substantive tensions with
models which assume extensions of the gauge and Higgs sectors with
breaking scales at the TeV range. Thus, the model can be tested
through the successes of the SM and the Dirac nature of light
neutrinos. Although we do not address here the phenomenon of dark
matter, we note that our model can be extended to accommodate its
particles, for instance, by adding extra sterile neutrinos.

Yet, the model maintains the expectations of the high scale type-I
seesaw mechanism.\cite{Minkowski}$^{\mbox{--}}$\cite{Moha} This
includes the new physics to be introduced with its energy
threshold ${\cal O}(10^{12} \, \mbox{GeV})$, required to solving
the problem of naturalness generated by the quantum corrections to
the Higgs boson mass. In fact, the type-I seesaw is natural up to
$M_{R} = {\cal O}(10^{7} \, \mbox{GeV})$,\cite{Vissani} but
demands an unnatural fine-tuning cancellation for $M_{R} = {\cal
O}(10^{12} \, \mbox{GeV})$. The only best known way of persisting
is through supersymmetry. This can be implemented in stages,
starting with the non-supersymmetric low-energy
theory,\cite{Urbano} although the high seesaw threshold calls for
the supersymmetrization of the Higgs, electroweak and fermionic
sectors. The no observation of supersymmetric phenomena may imply
that the quantum corrections to the Higgs mass would be suppressed
by another kind of new physics.

On the other hand, the model is crucially based on the hypothesis
of a symmetry of quark and lepton contents, which, however, is
validated by presymmetry. Light Dirac neutrinos are then
predictions of presymmetry in the SM extended with two RH
neutrinos per generation, implemented with a high scale seesaw
mechanism. Moreover, presymmetry of the SM with adulterated Dirac
neutrinos appears as a residue after removal of the heavy RH
neutrinos of Majorana type. Thus, the signatures of presymmetry
are also marks of the proposed SM with adulterated Dirac
neutrinos. Besides light Dirac neutrinos, they include
explanations of the fractional charge of quarks and quark--lepton
charge relations, understanding of the equality between the number
of generations and the number of quark colors, accounting for the
topological charge conservation in quantum flavor dynamics, and
elucidation of the charge quantization and the no observation of
fractionally charged hadrons.\cite{EAM1,EAM2}

\section{Conclusions}
\label{conclusions}

In the scenario of the SM extended with one RH neutrino per
generation, a simple paradigm for understanding the small value of
neutrino masses compared to the charged leptons is the type-I
seesaw mechanism where RH neutrinos have Majorana masses in
addition to Dirac masses. A distinguishing feature of this
mechanism is the Majorana nature of light neutrinos, which,
however, is not favored by recent experimental data on double-beta
decay of nuclei. Moreover, the current experimental situation
shows an agreement with the SM predictions well above the TeV,
apart from the presumed Dirac neutrinos with small masses for
which it gives no explanation.

Assuming no serious departures from the SM expectations at the TeV
range and the Dirac character of light neutrinos, recently we
proposed an extended seesaw in which two RH neutrinos per
generation are added, implemented with the hypothesis of the
symmetry of lepton and quark contents in order to restrain the
number of RH neutrinos from freedom, produce Dirac neutrinos and
naturally give them tiny masses. The first one is the usual RH
neutrino which re-establishes the correspondence between quarks
and leptons at high energies with weak couplings having order of
magnitudes as those of its weak charged partner and a Majorana
mass term whose coupling is assumed to be large, as in the
canonical high-scale type-I seesaw scenario. The second RH
neutrino, which breaks the quark--lepton symmetry founded with the
first one, has small masses and couplings, as explained by the
't~Hooft's naturalness criterion applied to this symmetry of
contents. The first RH neutrino is decoupled at the high scale,
while the second one survives down to the low scale to pair off in
a Dirac-like fashion with the corresponding LH neutrino, driving
its pattern of the symmetry of fermionic content. These symmetries
of particle content were only regarded as guidelines to the choice
of parameters since they cannot be understood as symmetry
transformations that exchange lepton and quark fields in the
Lagrangian of the model. It was supposed, however, that the proper
symmetry to take on has to be hidden in the SM with RH neutrinos
itself.

From another viewpoint, presymmetry was assumed as a symmetry that
underlies the SM extended with three RH neutrinos having Dirac
mass terms, so restoring lepton--quark symmetry of particle
content, unifying the electroweak properties of leptons and
quarks, and explaining the observed charge relations and chiral
structure of weak interactions. However, this scenario cannot
account naturally for the smallness of Dirac neutrino mass terms
relative to those of charged leptons. Besides, if it is pointed
out that the $B-L$ symmetry is an accidental symmetry of the SM
and that presymmetry is a hidden electroweak symmetry at a bare
level, the inclusion of Majorana mass terms for neutrinos seems
natural. Thus, both Dirac and Majorana mass terms are included in
the minimal SM extension of only three RH neutrinos, explicitly
breaking the conservation of the lepton number. Presymmetry still
appears as a basic symmetry of the model with neutrinos having
generic mass terms, since it is an extra enhanced symmetry
established at the bare step; such presymmetry transformations
perform at the underlying level of topological bare states which
have the same electroweak charges and no Majorana mass terms. Yet,
the tiny mass of Dirac-like neutrinos respect to charged leptons
still remains unnatural, so that the addition of just three RH
neutrinos with generic masses to the SM is not enough for
understanding such smallness.

The addition of a second RH neutrino per generation, however,
provides the seed for the expected light Dirac neutrinos. We have
shown that presymmetry defines properly the symmetry
transformations required by the symmetry of lepton and quark
contents and the assumed sequence of hierarchies. This means that
the extra Dirac and Majorana masses are not free parameters,
independent of the first RH neutrinos. We emphasize that these are
not made small by fine-tuning. Their smallness compared to the
first RH neutrinos are guaranteed by the presymmetry defined with
the usual RH neutrinos, at the high-energy seesaw scale. The
't~Hooft's argument of naturalness for the small values of the
Dirac and Majorana mass terms of the extra RH neutrinos in the
Lagrangian relies on this presymmetry with the first RH neutrinos;
as the couplings of the extra RH neutrinos tend to zero, the
underlying theory only involving the first RH neutrinos becomes
more symmetric. This symmetry guarantees the quantum corrections
of such parameters to be proportional to the parameters themselves
and its interplay with the seesaw mechanism leading to the
low-energy effective theory with the original RH neutrinos
decoupled only introduces negligible corrections to the mass
parameters. In particular, the smallness of the Dirac mass of
neutrinos compared to the charged leptons is stable.

Now a low scale Dirac scenario with symmetry of particle content
and small neutrino masses appears natural, satisfying 't~Hooft's
naturalness conditions. But, they involve the additional and not
the standard RH neutrinos, which are decoupled. The claim is that
the SM extended with extra RH neutrinos and implemented with the
seesaw mechanism and presymmetry, or the presymmetry model
implemented with extra RH neutrinos and the seesaw mechanism,
makes natural the existence of light Dirac-like neutrinos.
Neutrinos with extremely small masses are predicted to be of
adulterated Dirac nature in the sense that the ordinary RH
components are replaced by the almost inert extra ones. Besides,
the parameter region considered in this approach makes irrelevant
to low energy processes the perturbation of the seesaw mechanism
on a description given in terms of light Dirac neutrinos,
foreseeing that experiments will not have sensitiveness to the
Majorana character of neutrinos predicted by the seesaw mechanism,
as in the case of the neutrinoless double-beta decay.

On the other hand, the signatures of presymmetry are also
noticeable features of the model of Dirac neutrinos, such as the
topological character of fractional charges and the relationship
between the number of generations and the number of quark colors.
Although there are no hard predictions for the masses and mixing
of light neutrinos, the model does provide a new line of physics
beyond the SM for exploration.

Nevertheless, the high energy seesaw threshold established by our
approach raises the issue on the naturalness of the
renormalization of the Higgs mass due to the quantum corrections
introduced by the new physics states associated with the seesaw.
The best known way to solve it is via supersymmetry. An eventual
tension between experimental data and supersymmetry expectations
at the TeV range to be tested by the LHC may imply that the
solution to this naturalness problem would be given by another,
still unknown new physics.

\section*{Acknowledgment}

This work was supported by Vicerrector\'{\i}a de Investigaci\'on,
Desarrollo e Innovaci\'on, Universidad de Santiago de Chile,
Usach, Proyecto DICYT No. 041431MC.

\end{document}